\documentstyle[aps,epsf,floats]{revtex}
\draft
\begin{document}

\title{Photoluminescence of the Incompressible Laughlin Liquid:\\
       Excitons, Charged Excitons, and Fractionally Charged Excitons}

\author{
   Arkadiusz W\'ojs}
\address{
   Department of Physics, 
   University of Tennessee, 
   Knoxville, Tennessee 37996, USA, \\
   and Institute of Physics, 
   Wroc{\l}aw University of Technology, 
   50-370 Wroc{\l}aw, Poland}
\maketitle

\begin{abstract}
\par\noindent
The photoluminescence (PL) of a two-dimensional electron gas 
(2DEG) in a high magnetic field is studied as a function of 
the filling factor and the separation $d$ between the electron 
layer and the valence hole.
Depending on the magnitude of $d$ relative to the magnetic 
length $\lambda$, two distinct regimes in the response of 
the 2DEG to the valence hole occur, with different elementary 
emission processes contributing to the PL spectrum.
At $d<\lambda$ (``strong coupling'' regime), the hole binds one 
or two electrons to form an exciton ($X$) or one of three possible 
charged exciton ($X^-$) states, a spin-singlet or one of two 
spin-triplets.
At $d>\lambda$ (``weak coupling'' regime), the hole decouples 
or binds one or two Laughlin quasielectrons to form fractionally 
charged excitons (FCX's).
The binding energies as well as the emission energies and 
intensities of all $X^-$ and FCX states are calculated.
\end{abstract}

\section{Introduction}
%%%%%%%%%%%%%%%%%%%%%%%%%%%%%%%%%%%%%%%%%%%%%%%%%%%%%%%%%%%%%%%%%%%%%%
The magneto-optical properties of a two-dimensional electron gas 
(2DEG) systems have been intensively studied experimentally\cite{%
heiman,turberfield,goldberg,kheng,buhmann,shields,finkelstein,%
glasberg,priest,hayne,munteanu,vanhoucke,yusa} and theoretically
\cite{dzyubenko1,macdonald,chen,stebe,x-dot,palacios,whittaker,%
x-fqhe,x-cf,x-tb,riva}.
It is known that in a dilute system confined in a so-called symmetric
quantum well (QW), the photoluminescence (PL) spectrum is determined 
by a charged-exciton complex $X^-$ (bound state of two electrons $e$ 
and a valence hole $h$) and its interaction with the remaining electrons.
The existence of $X^-$ in bulk semiconductors was predicted by Lampert
\cite{lampert}, but only in 2D quantum wells (QW's) does its binding 
energy turn out sufficiently large\cite{stebe} for the experimental 
detection\cite{kheng,buhmann,shields,finkelstein,glasberg,priest,%
hayne,munteanu,vanhoucke,yusa}.
The observation of the $X^-$ stimulated theoretical work\cite{chen,%
stebe,x-dot,palacios,whittaker,x-fqhe,x-cf,x-tb,riva}, and it is now 
established that the only bound $X^-$ state at zero magnetic field is 
the singlet state ($X^-_s$) with the total electron spin $J=0$.
Accordingly, the PL spectrum usually shows two peaks, due to the $X$ 
(neutral exciton) and/or $X^-_s$ recombination, split by the $X^-_s$ 
binding energy $\Delta_s$.

The situation is more complicated in a magnetic field $B$. 
At very high $B$, the optically active $X$'s decouple from the electrons 
due to the ``hidden symmetry''\cite{dzyubenko1,macdonald,chen} and the 
$X^-_s$ unbinds. 
Interestingly, a different $X^-$ state binds in high fields\cite{x-dot}.
It is a triplet ($X^-_t$) with $J=1$ and finite total angular momentum.
It has infinite radiative lifetime $\tau_t$\cite{palacios} because of 
(independently) the ``hidden'' and 2D translational symmetry, and thus
will be further called $X^-_{td}$ ($d$ for ``dark'').
Although both symmetries are broken in experimental systems by mixing 
of Landau levels (LL's), valence band mixing effects, asymmetry of the 
QW, and disorder, the $X^-_{td}$ recombination is expected to be weak
and disappear at very large $B$.

The fact that $X^-_{td}$ unbinds at $B=0$ while $X^-_s$ unbinds at high 
$B$ implies a singlet--triplet crossing, which in a GaAs QW of width 
$w=10$~nm was predicted at $B\approx30$~T\cite{whittaker}.
Surprisingly, the following PL experiment\cite{hayne} showed no 
such transition up to $B\approx50$~T.
This puzzle was resolved by the identification of another bound state, 
a radiative excited spin-triplet $X^-_{tb}$ ($b$ for ``bright'')
\cite{x-tb}.
The emission spectrum from both triplets was eventually measured
\cite{yusa}, and agrees well with the theoretical prediction\cite{x-tb}.

The fact that the PL spectrum of a 2DEG in a symmetric QW only 
measures the $X$ and/or $X^-$ emission means that it is not a 
useful probe of electron correlations.
The effect of surrounding 2DEG on the $X^-$ emission is specially 
weak in dilute systems due to the Laughlin correlations effectively 
isolating an $X^-$ from the electrons\cite{x-fqhe,x-cf,x-tb}.
Indeed, the measured PL spectra are remarkably insensitive to the 
electron density\cite{priest,yusa}, and only at filling factors 
$\nu$ approaching ${1\over3}$ does the relative position of 
PL peaks change with density.

The PL spectra containing more information about the original 
(Laughlin) electron correlations of the 2DEG are obtained in 
asymmetrically doped wide QW's.
In such structures, the spatial separation $d$ of $e$ and $h$ 
layers weakens the $e$--$h$ interaction\cite{macdonald}, and 
the PL spectra show discontinuities\cite{heiman,turberfield,goldberg} 
at the filling factor $\nu={1\over3}$ at which Laughlin 
incompressible liquid state\cite{laughlin} is formed and the 
fractional quantum Hall (FQH) effect\cite{tsui} is observed in 
transport experiments.
The reason for these discontinuities is that at $d$ exceeding the 
magnetic length $\lambda$, the effective Coulomb potential of the 
hole seen by the electrons is too weak and its resolution is too low, 
and a hole can no longer bind ``whole'' electrons to form $X^-$'s.
Instead, the hole interacts with charge excitations that are already 
present in the 2DEG and which near $\nu={1\over3}$ happen to be 
a small number of Laughlin quasiparticles (QP's).
At $\nu<{1\over3}$, the hole repels positively charged quasiholes 
(QH's) and thus remains in the locally undisturbed incompressible 
electron state (such state can be called a ``decoupled hole'' and 
denoted by $h$)\cite{fcx}.
The situation is quite different at $\nu>{1\over3}$, when the hole 
attracts negatively charged quasielectrons (QE's) and binds one or 
two of them to form a fractionally charged exciton (FCX), $h$QE or 
$h$QE$_2$\cite{fcx}.

In this paper, we discuss results of numerical calculations of various 
radiative $e$--$h$ complexes that contribute to the PL spectrum 
of a 2DEG depending on $d$ and $\nu$.
The energy and PL spectra are obtained from exact diagonalization in 
Haldane spherical geometry\cite{haldane1,fano}.
The presented unified description of the PL from the 2DEG at high $B$
given in terms of emission from well defined bound states ($X$, $X^-$, 
and $h$QE$_n$) reduces the problem of a very complicated interaction
between a hole and the 2DEG to a simpler one, of determining the
single-particle properties of the bound ``quasiparticles''.
In particular, the optical selection rules for these states are 
formulated, following from the translational invariance.

\section{Model}
%%%%%%%%%%%%%%%%%%%%%%%%%%%%%%%%%%%%%%%%%%%%%%%%%%%%%%%%%%%%%%%%%%%%%%
Preserving the 2D translational symmetry of an infinite 2DEG in a 
finite-size calculation turns out to be essential for the identification 
of the optical selection rules of bound $e$--$h$ states and the 
degeneracies in their energy spectrum.
For any system of electrons and/or holes, this symmetry causes 
conservation of two orbital quantum numbers.
For systems with total electric charge, ${\cal Q}\ne0$ (such as $X^-$),
they are the total angular momentum ${\cal M}$ and an additional angular 
momentum quantum number ${\cal K}$ associated with a partial decoupling 
of the CM motion in a homogeneous magnetic field\cite{avron,dzyubenko2}.
The energy levels of a charged system fall into degenerate LL's labeled 
by ${\cal L}={\cal M}+{\cal K}$, and the states within each LL have 
${\cal K}=0$, 1, 2, \dots.
Since both ${\cal M}$ and ${\cal K}$ commute with the PL operator
${\cal P}$, the following selection rule governs recombination of 
isolated charged $e$--$h$ complexes: $\Delta{\cal M}=\Delta{\cal K}=0$.

In order to confine electrons to a finite area (in order to achieve 
finite degeneracy of electron and hole LL's) without breaking the 2D 
symmetry, Haldane \cite{haldane1} proposed to put electrons and holes 
on a sphere of radius $R$.
The magnetic field $B$ perpendicular to the surface is due to a monopole 
placed in the center.
The monopole strength $2S$ is defined in the units of elementary flux
$\phi_0=hc/e$, so that $4\pi R^2B=2S\phi_0$ and the magnetic length 
is $\lambda=R/\sqrt{S}$.
The single-particle states are the eigenstates of angular momentum 
$l$ and its projection $m$ and are called monopole harmonics
\cite{haldane1,fano}.
The energies $\varepsilon$ fall into $(2l+1)$-fold degenerate angular 
momentum shells separated by the cyclotron energy $\hbar\omega_c$.
The $n$-th ($n\ge0$) shell (LL) has $l=S+n$ and thus $2S$ is a measure 
of the system size through the LL degeneracy.
Due to the spin degeneracy, each shell is further split by the Zeeman 
gap, $E_Z$.

As a result of 2D rotational invariance, a many-body $e$--$h$ 
system on a sphere has two good quantum numbers, length $L$ and 
projection $L_z$ of the total angular momentum ${\bf L}$.
The mapping between quantum numbers ${\cal M}$ and ${\cal K}$ 
on a plane and the 2D algebra of ${\bf L}$ on a sphere allows 
conversion of the results between the two geometries\cite{fano}.
In particular, LL's of a charged $e$--$h$ complex are represented
on a sphere by $L$-multiplets, states within each LL are labeled 
by $L_z$, and the optical selection rule is $\Delta L=\Delta L_z=0$.
The price paid for closing the many-body Hilbert space without 
breaking the 2D symmetry is the surface curvature affecting
interaction matrix elements.
However, if the correlations modeled have short range $\xi$, the 
effects of curvature (scaled by a small parameter $\xi/R$) can be 
eliminated by extrapolation to $R\rightarrow\infty$.

Using a composite index $i=[nm\sigma]$ ($\sigma$ is the spin 
projection), the $e$--$h$ Hamiltonian can be written as
$H=\sum c_{i\alpha}^\dagger c_{i\alpha} \varepsilon_{i\alpha}
+\sum c_{i\alpha}^\dagger c_{j\beta}^\dagger c_{k\beta} c_{l\alpha} 
V^{\alpha\beta}_{ijkl}$, where $c_{i\alpha}^\dagger$ and $c_{i\alpha}$ 
create and annihilate particle $\alpha$ ($e$ or $h$) in state $i$, and 
$V^{\alpha\beta}_{ijkl}$ are the Coulomb matrix elements.
The Hamiltonian $H$ is diagonalized numerically in the basis of 
$Ne$--$1h$ Slater determinants.
Small density of holes is assumed which allows ignoring $h$--$h$ 
interaction effects and inclusion of only one valence hole in the 
basis.
The $Ne$--$1h$ eigenstates are labeled by $L$, $L_z$, and $J$.

The most important factors that determine dimension of the Hamiltonian 
matrix that must be diagonalized are the number of electrons $N$, the 
LL degeneracy controlled by $2S$, the number of excited LL's included, 
and the inclusion of spin-unpolarized states.
Using modified Lanczos algorithms we diagonalized Hamiltonians of 
dimensions up to about $5\times10^6$.
In the $X^-$ problem, the number of particles is small and the 
inclusion of reversed spins and LL's is possible.
The numerical results for $X^-$'s agree well with recent experiments
\cite{hayne,vanhoucke,yusa}.
However, when studying FCX's, as many electrons as possible need be 
included in the calculation, and the model must be simplified by 
assuming maximum spin polarization ($J={1\over2}N$), neglecting 
excited LL's, and setting zero QW width, $w=0$.
This allowed for calculation for $N\le9$ at $\nu\sim{1\over3}$, 
but the results are largely qualitative.

\clearpage
\section{Coulomb Potential of the Hole as a Perturbation:\\
         Strong and Weak Coupling Regimes}
%%%%%%%%%%%%%%%%%%%%%%%%%%%%%%%%%%%%%%%%%%%%%%%%%%%%%%%%%%%%%%%%%%%%%%

The potential $V_{UD}(r)$ seen by the electrons due to the positive 
charge of the optically injected valence hole $h$ can be described by 
two effective parameters, strength $U$ and spatial resolution $D^{-1}$.
The response of the 2DEG to the perturbation $V_{UD}(r)$ depends on
the relation between $U$ and $D$ and the characteristic energies and 
lengths of the unperturbed system.
The pair of perturbation parameters $U$ and $D^{-1}$ in principle 
can be varied independently, but in experimental samples they are
predominantly controlled through the effective spatial separation 
$d$ of the hole from the 2DEG layer.
The magnitude of $d$ depends on the electric field oriented across 
the 2DEG plane and resulting from asymmetric doping.
In symmetrically doped QW's, the hole moves in the same physical 
layer as the electrons ($d=0$).
However, in the asymmetrically (one-sided) doped QW's the displacement
$d$ can be as large as the QW width $w$.

Clearly, at $d=0$ both $U$ and $D^{-1}$ are the largest, and the
strongest response of the 2DEG to the hole can be expected.
The actual magnitudes of $U$ and $D^{-1}$ at $d=0$ depend on the
details of hole wave function, which to some extent depend on the
magnetic field.
In the extreme case of $B=\infty$ (lowest LL approximation), the 
$e$ and $h$ wave functions are identical and the interaction 
Hamiltonian is particle--hole symmetric\cite{dzyubenko1,macdonald,chen}.
This ``hidden'' symmetry causes vanishing of all electric moments 
in the optically active ground state of a bound $e$--$h$ pair ($X$), 
which in turn causes the decoupling of $X$'s from the 2DEG.
As a result, the $X$ is the most strongly bound optically active
$e$--$h$ state that can form in the 2DEG.
In the opposite case, the (infinitely heavy) hole acts as a point 
charge and its attraction to an electron is larger by a factor of 
$\sim\sqrt{2}$.
In an intermediate situation, the enhanced $e$--$h$ attraction 
always breaks the hidden symmetry and causes coupling of an $X$ to 
electrons.
As a result, radiative $X^-$ states occur, and the $X^-$ rather 
than $X$ should be regarded the most stable bound state formed by 
a hole injected into the 2DEG\cite{x-tb}.
This can be rephrased as that at finite $B$ and $d=0$ the 2DEG responds
to $V_{UD}(r)$ by binding two electrons to the hole to form an $X^-$
and screen the hole's positive charge.

In the other limit of large $d$, both $U$ and $D^{-1}$ are too small 
to cause a strong response of an incompressible electron liquid.
A hole is no longer able to pick out and bind a single electron from 
the 2DEG, which can be understood by noticing that the elementary 
charge excitations of an unperturbed 2DEG (QE's and QH's) have finite 
energy ($\varepsilon_{\rm QE}$ and $\varepsilon_{\rm QH}$), and that 
the electron wave function in the vicinity of the hole that corresponds
to an $X^-$ can be expanded in the basis of these QE and QH excitations.
On the other hand, too large $d$ implies too large $D$ and the size 
of an isolated $X^-$ ($2e$--$1h$ ground state) that would exceed the 
characteristic $e$--$e$ distance $\sqrt{\varrho}$ ($\varrho$ means 
density) and make such an $X^-$ unstable when inserted into the 2DEG 
(independently of the preceding energetic argument).
The only allowed response of the 2DEG to the hole in this (``weak'') 
coupling regime is to screen the hole's positive charge with the 
negatively charged QE's that are already present in the 2DEG.
The fact that only the existing QE's can bind to the hole makes the 
response critically depend on the filling factor $\nu$.
For example, near the Laughlin filling of $\nu={1\over3}$, the hole 
repels the QH's and moves in the locally incompressible $\nu={1\over3}$ 
liquid, causing no local response of the 2DEG.
Conversely, the binding of one or more QE's to the hole is expected at 
$\nu>{1\over3}$, and the resulting bound FCX states $h$QE$_n$ are the 
most stable bound states at large $d$.
Since the ``uncoupled'' state $h$ has different emission energy 
$\hbar\omega$ and oscillator strength $\tau^{-1}$ than the $h$QE$_n$'s, 
discontinuity in the PL spectrum is expected at $\nu={1\over3}$, 
in agreement with the experimental PL data for asymmetric structures
\cite{heiman,turberfield,goldberg}.

\section{Strong Coupling Regime: Neutral and Charged Excitons}
%%%%%%%%%%%%%%%%%%%%%%%%%%%%%%%%%%%%%%%%%%%%%%%%%%%%%%%%%%%%%%%%%%%%%%

\subparagraph*{Isolated Charged Exciton: Energy Spectrum.}
% % % % % % % % % % % % %% % % % % % % % %% % % % % % % % % % % % % % 
An isolated neutral ($X$) or charged ($X^-$) exciton consists of 
only two or three particles.
In the numerical calculation of the energy and PL spectra of this
relatively simple quantum mechanical system we were able to include 
the effects of the finite magnetic field $B$ that causes LL mixing
(by including up to five electron and hole LL's, $n\le4$, with the 
lowest LL degeneracy of $2S+1=21$) and of the finite QW width $w$
(by using the values of the two-body Coulomb matrix elements appropriate 
for the lowest subband of the QW rather than for an ideal 2D system).
It should be stressed that due to different electron and hole effective 
masses and due to a different height of the electron and hole confinement 
potential at the well/barrier interface, the electron and hole wave 
functions in the $z$-direction are different even in the symmetric 
structures with $d=0$.
To a first approximation, this is included by introducing a pair of
effective widths of the electron and hole layers, $w_e^*$ and $w_h^*$,
obtained by fitting the actual density profile $\varrho(z)$ of each 
carrier with $\cos^2(\pi z/w^*)$.
For a $w=10$~nm GaAs/Al$_{0.33}$Ga$_{0.67}$As QW the effective widths
are $w_e^*=w+3.3$~nm and $w_h^*=w+1.5$~nm. 

The Zeeman energy $E_Z=g^*\mu_B B$ ($g^*$ is the effective giromagnetic 
factor and $\mu_B$ is the Bohr magneton) enters the problem of an $X^-$ 
in two ways.
(i) Because of the complete spin polarization of the 2DEG, binding of 
an $X^-$ state may or may not require a spin flip, depending on the 
total spin $J$ of two electrons in the bound $X^-$ state.
Therefore, the binding energy $\Delta$ of each spin triplet ($J=1$) 
$X^-$ state is insensitive to the Zeeman energy, but for spin singlet 
($J=0$) states $\Delta$ is reduced by $E_{Ze}$ compared to the pure 
Coulomb binding energy.
For GaAs, $E_{Ze}$ is roughly a linear function of energy through 
both cyclotron energy $\hbar\omega_c\propto B$ and confinement 
energy $\propto1/w^2$.
After Snelling et al.\cite{snelling}, for $w\sim10$~nm at $B=0$, 
we have $g_e^*=(w_0/w)^2+g_0$ with $w_0=9.4$~nm and $g_0=-0.29$.
After Seck et al.\cite{seck} we find $dg_e^*/dB=0.0052$~T$^{-1}$.
(ii) The recombination of an $X$ or $X^-$ state can occur through 
relaxation of a conduction electron with either spin to the valence 
band.
Since the angular momentum of the photon ($\pm1$) depends on the 
spin of the relaxing electron, and the emission energy $\hbar\omega$ 
includes the Zeeman energy of the recombining $e$--$h$ pair, 
$E_{Ze}+E_{Zh}$, each PL peak associated with the emission from an 
$X$ or $X^-$ state is split into two circularly polarized peaks by 
$E_{Ze}+E_{Zh}$.
The splitting is most evident for $X^-_s$ because of the large
population of both electron spin levels even at a very low 
temperature.

\begin{figure}
\centering
\epsfxsize=4.8in
\epsffile{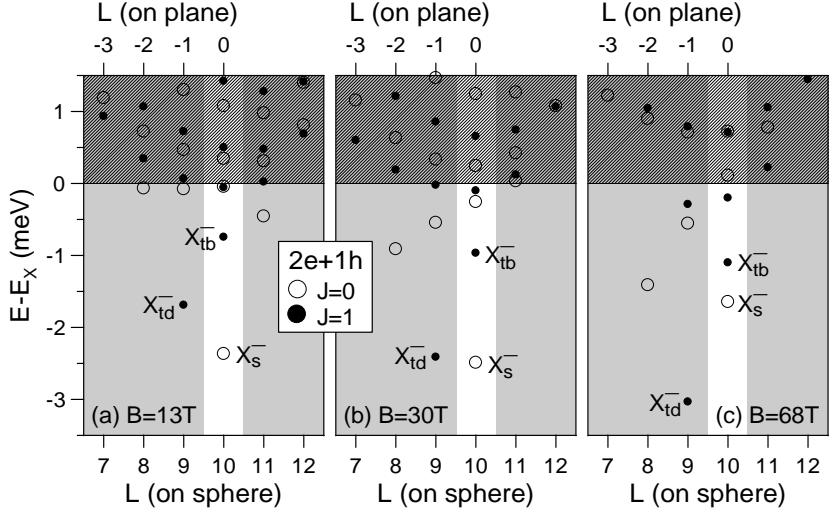}
\caption{
   The energy spectra (energy $E$ vs.\ angular momentum $L$) of 
   two electrons and one valence hole confined in a symmetric 
   GaAs quantum well of width $w=11.5$~nm at the magnetic field 
   $B=13$~T (a), 30~T (b), and 68~T (c), calculated on a Haldane 
   sphere with the Landau level degeneracy $2S+1=21$.
   $E_X$ is the exciton energy.
}
\label{fig1}
\end{figure}

In Fig.~\ref{fig1} we show the $2e$--$1h$ energy spectra calculated 
for a $w=11.5$~nm GaAs QW at $B=13$, 30, and 68~T.
In each frame, the energy $E$ is measured from the exciton energy 
$E_X$, so that for the bound $X^-$ states it is opposite to the 
binding energy $\Delta$.
Open and full symbols denote singlet and triplet electron spin 
configurations, respectively, and only the state with the lowest 
Zeeman energy ($J_z=J$) is marked for each triplet.
Similarly, each state with $L>0$ represents a degenerate multiplet 
with $|L_z|\le L$.
The angular momentum $L$ calculated in the spherical geometry and
given on the horizontal axes under each graph translates into the 
angular momentum on a plane ${\cal L}$ in such way\cite{avron,%
dzyubenko2} that the $L=S$ and $S-1$ multiplets correspond to the 
planar LL's with ${\cal L}=0$ and $-1$, respectively.

\subparagraph*{Isolated Charged Exciton: Radiative Recombination.}
% % % % % % % % % % % % %% % % % % % % % %% % % % % % % % % % % % % % 
Due to the conservation of $L$ in the PL process, only those $2e$--$1h$
states from the $L=S$ channel are radiative.
This is because\cite{chen,x-fqhe,x-cf,x-tb} an annihilated $e$--$h$ pair 
has $l_X=0$, and the electron left over in the lowest LL has $l_e=S$.
Recombination of other, non-radiative states requires breaking the 
rotational symmetry (e.g., by interaction with other charges).
This result is independent of chosen spherical geometry; on a plane 
the 2D translational symmetry leads to the conservation of both 
${\cal M}$ and ${\cal K}$, and the corresponding PL selection rule 
for $2e$--$1h$ states is ${\cal L}=0$\cite{dzyubenko2}.

\begin{figure}
\centering
\epsfxsize=4.8in
\epsffile{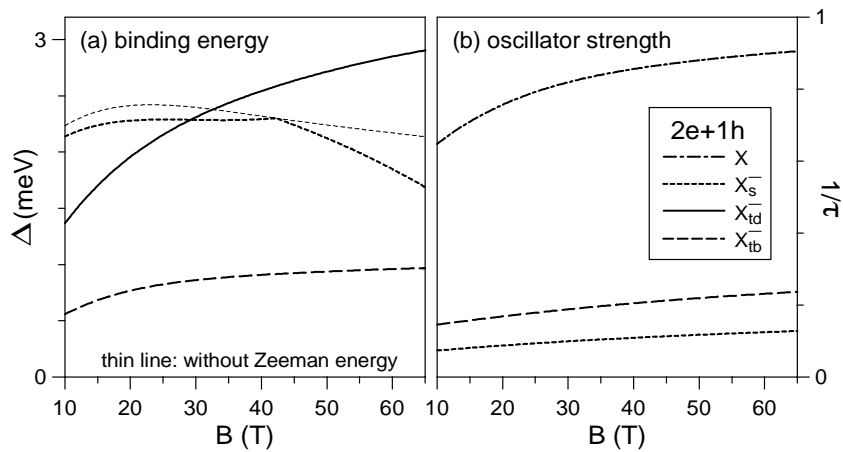}
\caption{
   The binding energies $\Delta$ (a) and oscillator strengths 
   $\tau^{-1}$ (b) of different charged exciton states $X^-$
   in a symmetric GaAs quantum well of width $w=11.5$~nm,
   plotted as a function of the magnetic field $B$.
}
\label{fig2}
\end{figure}

Three $X^-$ states in Fig.~\ref{fig1} are of particular importance.
The $X^-_s$ and $X^-_{tb}$, the lowest singlet and triplet states at 
$L=S$, are the only strongly bound radiative states, while $X^-_{td}$ 
has by far the lowest energy of all non-radiative ($L\ne S$) states.
In agreement with earlier prediction\cite{whittaker}, the transition 
from $X^-_s$ to $X^-_{td}$ ground state is found at $B\approx30$~T,
and a new, radiative excited triplet state $X^-_{tb}$ is identified
in all frames.
The binding energy $\Delta$ of each of these three $X^-$ states,
extrapolated to the $R/\lambda=\sqrt{S}=\infty$) limit, is plotted 
as a function of $B$ in Fig.~\ref{fig2}a.
Clearly, the $\Delta_{td}$ increases most quickly of all curves with 
increasing $B$, the $\Delta_s$ remains almost constant (specially its 
Coulomb part drawn with the thin line), and $\Delta_{tb}$ remains 
smaller than both $\Delta_{td}$ and $\Delta_s$ at any value of $B$.

The extrapolated values of the oscillator strength $\tau^{-1}$ of 
the $X$ and two radiative $X^-$ states are shown in Fig.~\ref{fig2}b.
The ratio $\tau_{tb}^{-1}\approx2\tau_s^{-1}$ remains almost 
independent of $B$, and the resulting three PL peaks ($X$, $X^-_s$, 
and $X^-_{tb}$) are precisely the ones observed in experiments\cite{%
shields,finkelstein,glasberg,priest,hayne,munteanu,vanhoucke,yusa}.
In the above, we assumed that both electrons and holes are completely 
spin-polarized ($J_z=J$), but typically, all electron spins and only 
a fraction of hole spins $\chi_h$ are aligned with the field.
As a result, the $X^-_{tb}$ PL has definite circular polarization 
($\sigma_+$) and its intensity is reduced by $\chi_h$, while the 
$X^-_s$ PL peak splits into a $\sigma_\pm$ doublet (separated by 
the appropriate Zeeman energy) with the intensity of the two 
transitions weighted by $\chi_h$ and $1-\chi_h$.
An increase of $\chi_h$ from ${1\over2}$ to 1 with increasing $B$ may
explain the observed\cite{hayne} increase of $\tau^{-1}_{tb}$ by up to 
a factor of two, while $\tau^{-1}_s$ remains constant.

\subparagraph*{Interaction of Charged Excitons with Electrons.}
% % % % % % % % % % % % %% % % % % % % % %% % % % % % % % % % % % % % 
Even in dilute systems, recombination of bound $e$--$h$ complexes can 
in principle be affected by their interaction with one another or with 
excess electrons.
Specially the recombination of a $X^-_{td}$ might become allowed in 
a collision assisted process in which the translational symmetry of
an isolated $X^-_{td}$ is broken.
The critical question is if the $e$--$X^-$ correlations are of 
the Laughlin type\cite{x-cf,laughlin}, meaning that the many-body 
$e$--$X^-$ wave function contains a Jastrow prefactor 
$\prod(x_i-y_j)^\mu$ (where $x$ and $y$ are complex coordinates of 
$e$ and $X^-$, and $\mu$ is an integer).
If it is so, then a number $\mu$ of the highest energy $e$--$X^-$ 
pair eigenstates are avoided in the low-energy many-body states
\cite{haldane2,parentage} (just as the $p$ leading $e$--$e$ pair 
eigenstates are avoided in the Laughlin $\nu=(2p+1)^{-1}$ state of 
electrons).
This means lack of high-energy $e$--$X^-$ collisions, and thus an 
effective isolation of the $X^-$ states from the 2DEG and insensitivity 
of the $X^-$ binding or recombination to the electron density.
In particular, Laughlin $e$--$X^-_{td}$ correlations would eliminate
the possibility of collision-assisted recombination of the $X^-_{td}$. 

To determine if the Laughlin correlations occur in a mixed $e$--$X^-$ 
liquid, one must calculate the $e$--$X^-$ interaction pseudopotential 
$V(L)$, defined as the dependence of the pair interaction energy $V$ 
on the pair angular momentum $L$\cite{haldane2}.
The general criterion for the occurrence of Laughlin correlations in 
a many-body system confined to a degenerate LL and interacting through 
$V(L)$ is that $V$ must have short range, i.e. decrease sufficiently 
quickly with increasing $L$\cite{parentage}.
On a sphere, $V$ must be a super-linear function of $L(L+1)$.
It turns out that this criterion is satisfied by the $e$--$X^-$ 
repulsion in narrow QW's\cite{x-tb}.
This implies a simple connection between $\nu$ and the maximum allowed 
$L$ (i.e., minimum average separation) for a $e$--$X^-$ pair:
$L+\mu\le l_e+l_{X^-}$ for $\nu\le\mu^{-1}$, and allows calculation 
of the effect of the $e$--$X^-$ interaction on the $X^-$ recombination 
as a function of $\nu$.

\begin{figure}[t]
\centering
\epsfxsize=4.8in
\epsffile{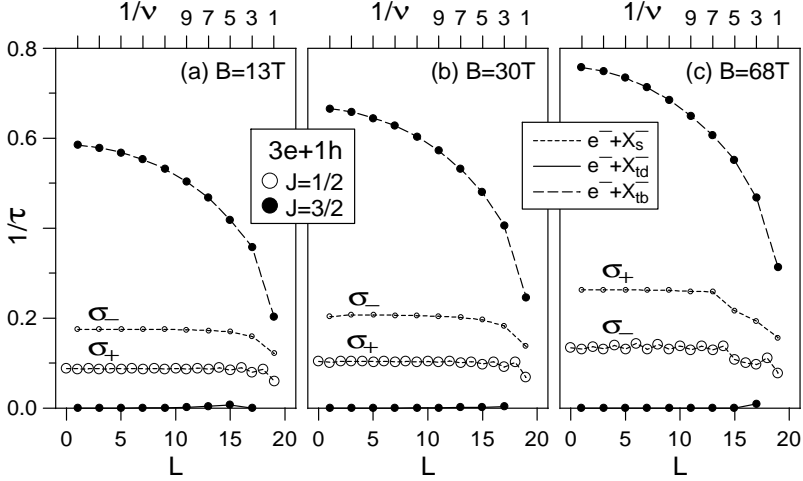}
\caption{
   The oscillator strengths $\tau^{-1}$ of different charged 
   excitons $X^-$ interacting with an electron in a symmetric 
   GaAs quantum well of width $w=11.5$~nm at the magnetic field 
   $B=13$~T (a), 30~T (b), and 68~T (c), calculated on a Haldane 
   sphere with the Landau level degeneracy $2S+1=21$, and plotted
   as a function of the $e$--$X^-$ pair angular momentum $L$.
}
\label{fig3}
\end{figure}

\begin{figure}[t]
\centering
\epsfxsize=4.9in
\epsffile{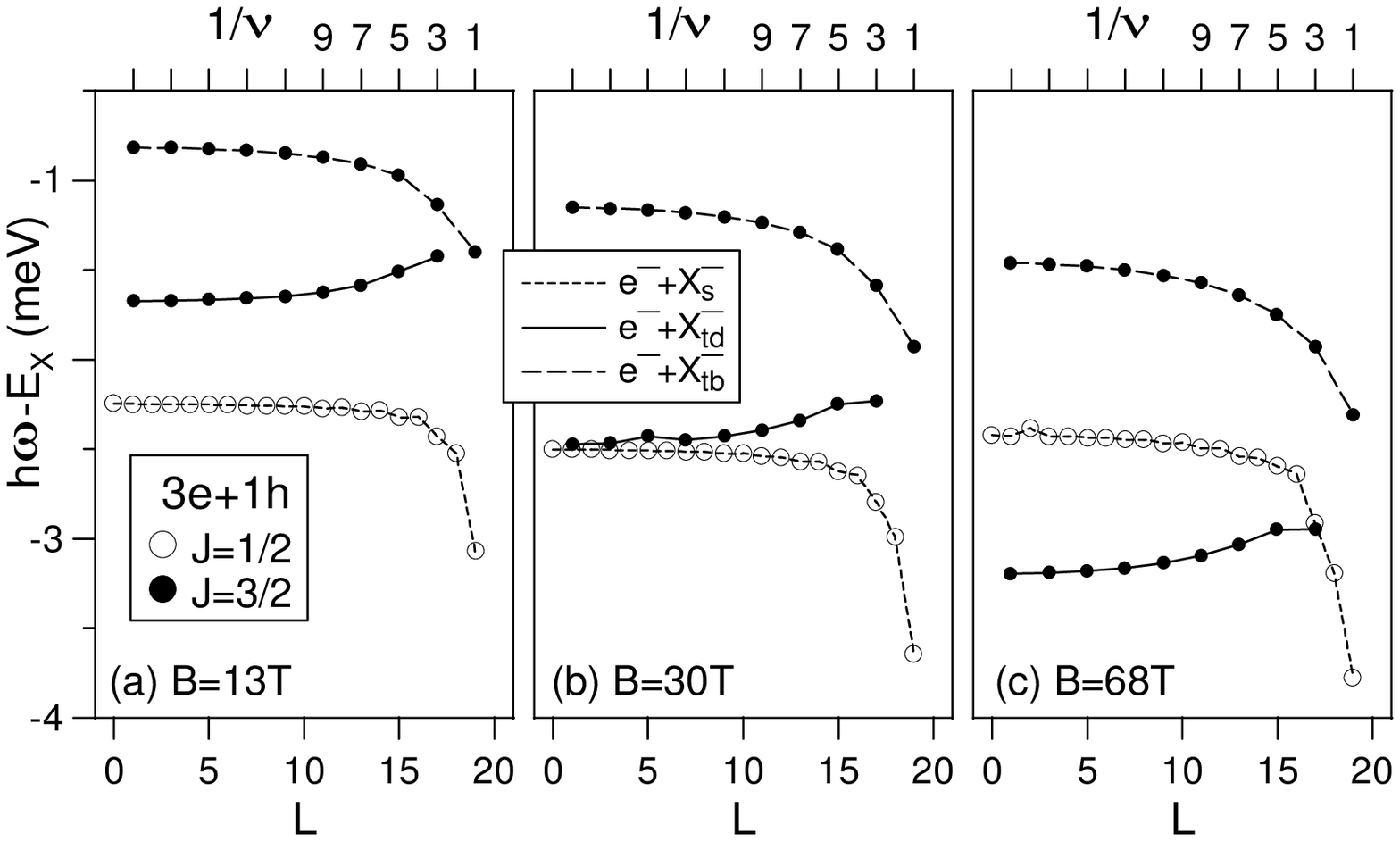}
\caption{
   The same as in Fig.~\ref{fig3} but the graphs show the 
   recombination energy $\hbar\omega$ as a function of $L$.
   $E_X$ is the exciton energy.
}
\label{fig4}
\end{figure}

In Figs.~\ref{fig3} and \ref{fig4} we plot the PL oscillator strength 
$\tau^{-1}$ and energy $\hbar\omega$ (measured from the exciton energy 
$E_X$) for the $3e$--$1h$ eigenstates corresponding to an $X^-$ 
interacting with another electron.
We assume that the Zeeman energy will polarize all electron spins prior 
to recombination, except for those two in the $X^-_s$, and concentrate 
on the following three initial configurations: 
$e$--$X^-_s$ with $J_z=J={1\over2}$ and
$e$--$X^-_{tb}$ and $e$--$X^-_{td}$ with $J_z=J={3\over2}$.
For each of the three configurations, $\tau^{-1}$ and $E$ are plotted 
as a function of $L$ (i.e.\ of $\nu$).

For $X^-_{tb}$ and $X^-_{td}$, only an $e\!\!\uparrow$--$h\!\!\downarrow$ 
pair can be annihilated, and an emitted photon has a definite circular 
polarization $\sigma_+$.
Two indistinguishable electrons left in the final state have $J=1$, 
so their $L$ must be odd ($2l_e$ minus an odd integer).
For $X^-_s$, both $\sigma_+$ and $\sigma_-$ PL are possible, with the
energy of the latter transition shifted by $E_{Ze}+E_{Zh}$.
For $\sigma_+$, the two electrons in the final state can have either 
$J=0$ and $L$ even, or $J=1$ and $L$ odd; while for $\sigma_+$ they 
can only have $J=1$ and $L$ must be odd.

As expected, for $L\rightarrow0$ (i.e., $\nu\rightarrow0$) both
$\hbar\omega$ and $\tau^{-1}$ converge to the values appropriate 
for single $X^-$'s plotted in Fig.~\ref{fig2}
(the energies shown in Fig.~\ref{fig4} correspond to the $\sigma_+$ 
transitions given by $E_X$ plus Coulomb binding energy; if present 
in the PL spectrum, additional $\sigma_-$ transitions will appear 
at the energy higher by $E_{Ze}+E_{Zh}$).
There is no significant effect of the $e$--$X^-$ interactions on the 
$X^-$ recombination at small $L$.
This justifies a simple picture of PL in dilute $e$--$h$ plasmas,
according to which, recombination occurs from a single isolated bound 
complex and hence is virtually insensitive\cite{priest} to $\nu$.
Somewhat surprisingly, the Laughlin correlations prevent increase 
of the $X^-_{td}$ oscillator strength through interaction with other 
charges ($\tau^{-1}_{td}$ remains ten times longer than $\tau_s$ even 
at $\nu={1\over3}$).
This explains the absence of an $X^-_{td}$ peak even in the PL spectra 
\cite{shields,finkelstein,glasberg,priest,hayne} showing strong 
recombination of a higher-energy triplet state $X^-_{tb}$ (except at
very low temperatures\cite{yusa}).
An interesting feature in Fig.~\ref{fig4} is the merging of 
$\hbar\omega_{tb}$ and $\hbar\omega_{td}$ which actually has
been observed\cite{yusa}.

\section{Weak Coupling Regime: Fractionally Charged Excitons}
%%%%%%%%%%%%%%%%%%%%%%%%%%%%%%%%%%%%%%%%%%%%%%%%%%%%%%%%%%%%%%%%%%%%%%

The fractionally charged excitons (FCX's) $h$QE$_n$ formed in strongly 
asymmetric QW structures ($d>\lambda$) consist of $n$ QE's bound to 
a valence hole $h$.
Since QE's are collective excitations of a many-electron system,
as many electrons as possible must be included in the computation.
This was only possible by a severe limitation of the single-particle
Hilbert space to the lowest LL.
Although the inter-LL $e$--$h$ scattering was quite important at $d=0$ 
(where it caused binding of $X^-_s$ and $X^-_{tb}$ states), it is 
greatly reduced at larger $d$ and, unlike at $d=0$, the lowest-LL 
approximation is expected to be more justified.
Nevertheless, the model studied is necessarily a very ideal one
and, consequently, such realistic elements as the finite QW width
have also been excluded.
Therefore, the numerical results obtained in this section are not 
meant to describe experiments as accurately as those for $X^-$'s,
although the fact of the FCX binding or the optical selection rules 
for different FCX states are expected to be valid for experimental
systems.

The binding of FCX states relies on the attractive interaction 
between the (oppositely charged) $h$ and QE, and the weak QE--QE 
interaction at short range\cite{hierarchy}.
On the other hand, the stability of the FCX states at sufficiently
large $d$ against the formation of $X$ or $X^-$ excitons results 
from the $h$--QE attraction being sufficiently small compared to the 
Laughlin gap $\varepsilon_L=\varepsilon_{\rm QE}+\varepsilon_{\rm QH}$.
The comparison of the $h$--QE attraction energy (the largest
$h$--QE pseudopotential parameter) to $\varepsilon_L$ shows that
the FCX's rather than $X$'s or $X^-$'s should be the most stable bound 
states formed by a hole injected into the 2DEG at $d$ larger than about
a magnetic length $\lambda$\cite{fcx}.

\begin{figure}
\centering
\epsfxsize=4.9in
\epsffile{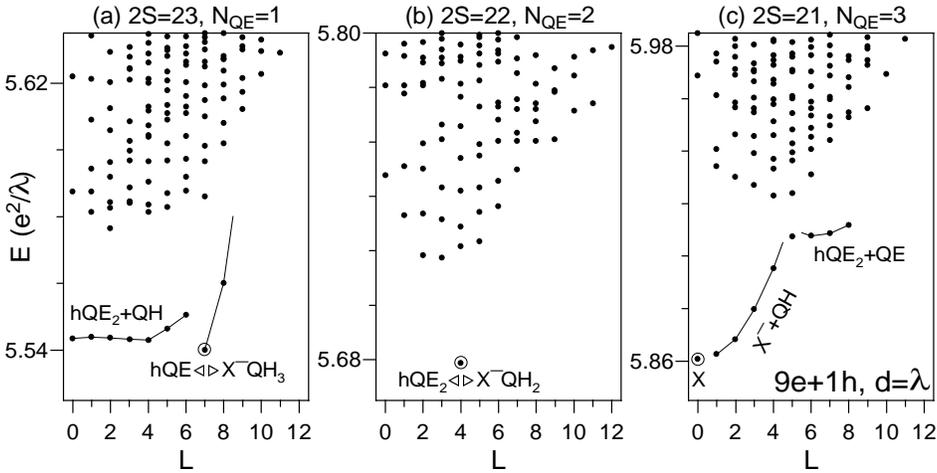}
\caption{
   The energy spectra (energy $E$ vs.\ angular momentum $L$) 
   of nine electrons and one valence hole on a Haldane sphere 
   with the Landau level degeneracy $2S+1=24$, 23, and 22,
   corresponding to $N_{\rm QE}=1$, 2, and 3 quasielectrons
   in the Laughlin $\nu={1\over3}$ state of nine electrons.
   The separation between electron and hole planes is $d=\lambda$.
   $\lambda$ is the magnetic length.
}
\label{fig5}
\end{figure}

\begin{figure}
\centering
\epsfxsize=4.9in
\epsffile{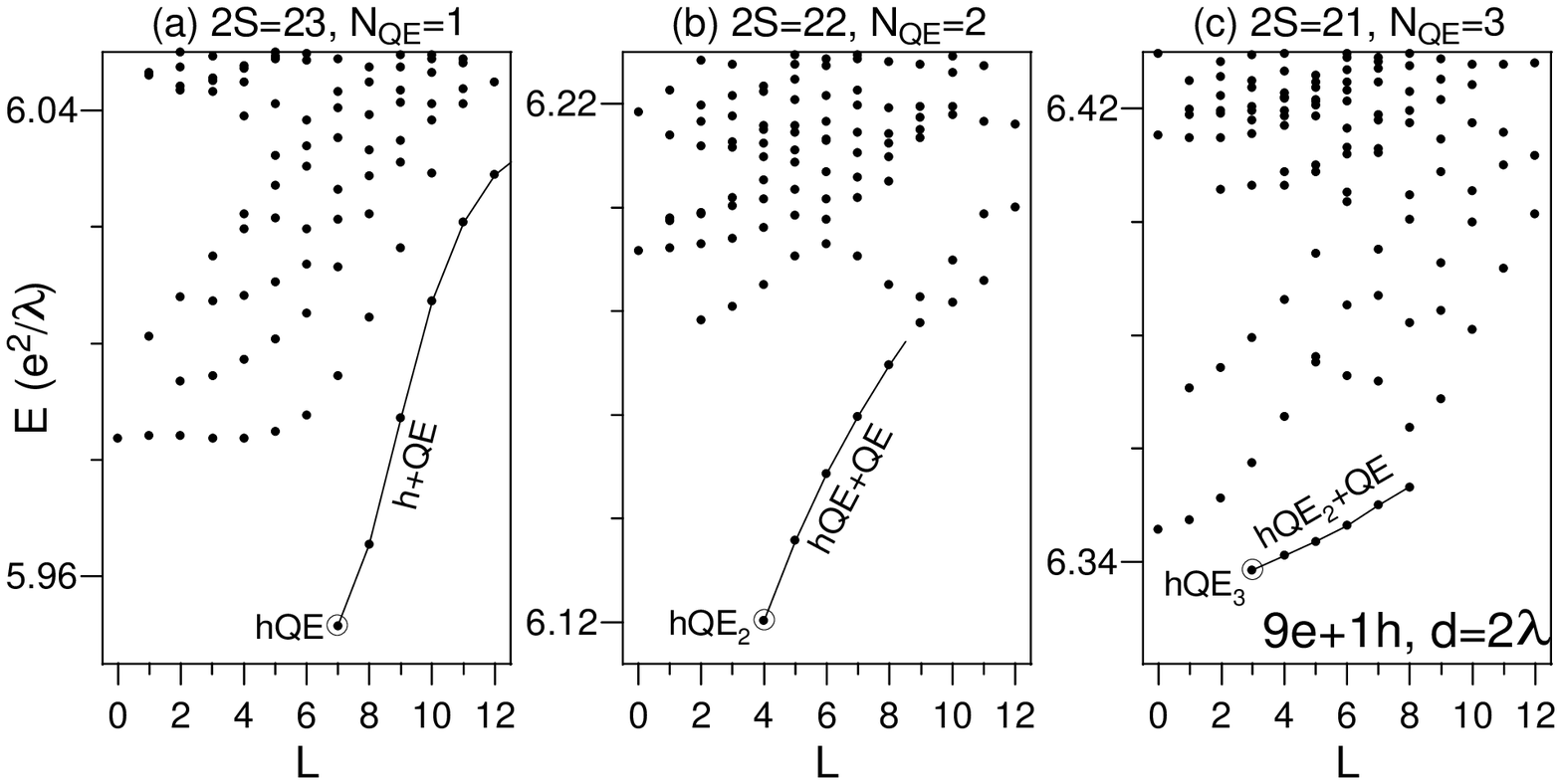}
\caption{
   The same as in Fig.~\ref{fig5} but for $d=2\lambda$.
}
\label{fig6}
\end{figure}

\subparagraph*{Numerical Energy Spectra.}
% % % % % % % % % % % % %% % % % % % % % %% % % % % % % % % % % % % % 
The series of $9e$--$1h$ energy spectra calculated for $d=\lambda$ and 
$2\lambda$ are presented in Figs.~\ref{fig5} and \ref{fig6}.
The three spectra shown for each $d$ are obtained for different values
of $2S$, corresponding to $N_{\rm QE}=1$, 2, and 3 QE's in the Laughlin 
$\nu={1\over3}$ state of $9e$ system (without interaction with the 
valence hole).
In Fig.~\ref{fig5} ($d=\lambda$, intermediate-coupling regime), new 
low-energy bands of states emerge in addition to those characteristic
of strong coupling and containing an $X$ or $X^-$.
These new states contain various FCX's interacting with the remaining 
QP's of the $9e$ liquid.
In some cases the FCX states occur in the same spectrum with the $X$ 
or $X^-$ states.
For example, the $h$QE$_2$--QE band in Fig.~\ref{fig5}c coexists with 
the $X$ state (meaning $X$ weakly coupled to the Laughlin $\nu={1\over3}$ 
state of the remaining $8e$) and the $X^-$--QH band (meaning $X^-_{td}$ 
Laughlin-correlated with the remaining $7e$; QH denotes a quasihole 
in the two-component, $e$--$X^-$ Laughlin liquid\cite{x-cf}).
In other cases, low-lying $X$ or $X^-$ states of the strong-coupling 
regime occur at the same $L$ as a low-lying FCX of the weak-coupling
regime, and the transition between the two states (which occurs at 
$d\approx\lambda$) is continuous.
For example, $h$QE$_2$ is mixed with $X^-$QH$_2$ in Fig.~\ref{fig5}b,
and $h$QE is mixed with $X^-$QH$_3$ in Fig.~\ref{fig5}a.

In Fig.~\ref{fig6} ($d=2\lambda$, weak-coupling regime), well developed 
FCX bands occur.
The isolated $h$QE, $h$QE$_2$, and $h$QE$_3$ states are the ground 
states in the spectra corresponding to $N_{\rm QE}=1$, 2, and 3, 
respectively.
Their angular momenta $l_{\rm FCX}$ are obtained by adding $l_h=S$ 
and $l_{\rm QE}=S^*+1$, where $2S^*=2S-2(N-1)$ is the effective monopole 
strength in the composite fermion picture\cite{chen,parentage} and 
$2S=3(N-1)-N_{\rm QE}$.
Similarly, the angular momenta of states containing an FCX and 
excess QP's result from adding $l_{\rm FCX}$ and $l_{\rm QP}$.

\subparagraph*{Selection Rules and Photoluminescence.}
% % % % % % % % % % % % %% % % % % % % % %% % % % % % % % % % % % % % 
Similarly as it was for $X^-$'s, the translational symmetry of an 
isolated FCX leads to the conservation of $L$ and $L_z$ in the 
emission process.
This leads to the strict optical selection rules that can only be 
broken by collisions or disorder.
The recombination of an FCX state $h$QE$_n$ formed in a Laughlin 
$\nu=(2p+1)^{-1}$ electron liquid occurs through annihilation of a 
well defined number of QE's and creation of an appropriate number
of QH's.
It turns out that the processes involving more than the minimum 
number of QP's all have negligible intensity, which for $p=1$ 
($\nu={1\over3}$) leaves only the following four possible 
recombination events: $h+n{\rm QE}\rightarrow (3-n){\rm QH}+\gamma$,
where $n=0$, 1, 2, or 3, and $\gamma$ denotes the photon.
Let us apply the angular momentum conservation law, $\Delta L
=\Delta L_z=0$, to the above recombination events.
In the fermionic picture\cite{hierarchy}, the angular momenta of 
QE and $h$ in the initial $Ne$--$1h$ state at a given monopole 
strength $2S$ are $l_{\rm QE}=S-N+2$ and $l_h=S$.
By adding $l_{\rm QE}$ and $l_h$, the following values are obtained
for the angular momenta of FCX complexes\cite{fcx}:
$l_{h{\rm QE}}=N-2$, $l_{h{\rm QE}_2}={1\over2}(N-1)$, and 
$l_{h{\rm QE}_3}=3$.
The angular momentum of QH in the final $(N-1)e$ state at the 
same $2S$ is $l_{\rm QH}=S-N+2$.
By comparing the values of $l_{h{\rm QE}_n}$ with the angular 
momenta allowed for $(3-n)$ identical QH's in the final state, 
we obtain that 
(i) the ``decoupled hole'' state $h$ is radiative and can recombine 
to create a QH$_3$ molecule with the maximum $L=3l_{\rm QH}-3$ 
allowed for three QH's, 
(ii) $l_{h{\rm QE}}$ is different from any $L$ allowed for two QH's
and thus $h$QE is non-radiative; however, the first excited state,
$h$QE* at $l_{h{\rm QE}^*}=l_{h{\rm QE}}+1$ is radiative and 
recombines to create a QH$_2$ molecule with $L=2l_{\rm QH}-1$;
(iii) $l_{h{\rm QE}_2}$ is radiative and its recombination leaves
behind a single QH;
(iv) neither $l_{h{\rm QE}_3}$ nor its excitations are radiative.

The above analysis leaves $h$, $h$QE*, and $h$QE$_2$ as the only
radiative FCX states, while $h$QE and $h$QE$_3$ are found dark.
It is expected that a valence hole introduced into the 2DEG at 
$\nu<{1\over3}$ (in the absence of free QE's) and at $d>\lambda$ 
will decouple and recombine from the initial state $h$ (local 
filling factor $\nu={1\over3}$).
On the other hand, at $\nu>{1\over3}$ the valence hole will bind 
one or more free QE's and recombine from either $h$QE* or $h$QE$_2$
initial state, depending on $d$, temperature, and the QE density.
Since the initial state from which the hole recombines changes at 
$\nu={1\over3}$, and since different FCX states have different
emission energy and intensity\cite{fcx}, the PL spectrum of a 2DEG
created in a strongly asymmetric structure is expected to change
discontinuously at the corresponding magnetic field.

\section{Conclusion}
%%%%%%%%%%%%%%%%%%%%%%%%%%%%%%%%%%%%%%%%%%%%%%%%%%%%%%%%%%%%%%%%%%%%%%
The PL spectrum of a 2DEG was studied as a function of the separation
$d$ between $e$ and $h$ layers.
Two types of response of the 2DEG to the optically injected hole
were identified.
In the strong-coupling regime ($d\ll\lambda$) the most strongly bound 
states are $X$ and $X^-$, and the PL spectrum measures their optical 
properties rather than the original correlations of the 2DEG.
In particular, the dark triplet $X^-_{td}$ remains virtually 
non-radiative at $\nu<{1\over3}$.
In the weak-coupling regime ($d>\lambda$), the $X^-$ states unbind 
and, depending on $\nu$, the hole either decouples from the 2DEG or 
binds one or more QE's to form a FCX state $h$QE$_n$.
The most stable radiative bound state at $\nu\le{1\over3}$ is found 
to be the ``decoupled hole,'' while at $\nu>{1\over3}$ it is either 
$h$QE$_2$ or $h$QE* (excited state of $h$QE).
Since different FCX states have different opticl properties, 
discontinuities are expected in the PL spectrum at $\nu={1\over3}$.

\subparagraph*{Acknowledgment}
% % % % % % % % % % % % %% % % % % % % % %% % % % % % % % % % % % % % 
The author acknowledges helpful discussions with J. J. Quinn, 
P. Hawrylak, M. Potemski, and I. Bar-Joseph, and support by the 
KBN grant 2P03B05518.

\end{document}